\documentclass[twocolumn]{aastex631}

\usepackage{url}
\usepackage{epsfig}
\usepackage{graphicx}
\usepackage{xcolor}
\usepackage{hyperref}
\usepackage{threeparttable}
\usepackage{multibib}
\usepackage{amssymb,amsmath}

\begin{document}

\title{Detection of hidden emissions in two rotating radio transients with high surface magnetic fields}

\email{jjgeng@pmo.ac.cn(JJG), ypyang@ynu.edu.cn(YPY), xfwu@pmo.ac.cn(XFW)}

\author{S.B. Zhang}
\affiliation{Purple Mountain Observatory, Chinese Academy of Sciences, Nanjing 210023, China}
\affiliation{CSIRO Space and Astronomy, Australia Telescope National Facility, PO Box 76, Epping, NSW 1710, Australia}

\author{X. Yang}
\affiliation{Purple Mountain Observatory, Chinese Academy of Sciences, Nanjing 210023, China}
\affiliation{School of Astronomy and Space Sciences, University of Science and Technology of China, Hefei 230026, China}
\affiliation{CSIRO Space and Astronomy, Australia Telescope National Facility, PO Box 76, Epping, NSW 1710, Australia}

\author{J.J Geng}
\affiliation{Purple Mountain Observatory, Chinese Academy of Sciences, Nanjing 210023, China}

\author{Y.P. Yang}
\affiliation{South-Western Institute For Astronomy Research, Yunnan University, Yunnan 650504, P. R. China}
\affiliation{Purple Mountain Observatory, Chinese Academy of Sciences, Nanjing 210023, China}

\author{X.F. Wu}
\affiliation{Purple Mountain Observatory, Chinese Academy of Sciences, Nanjing 210023, China}
\affiliation{School of Astronomy and Space Sciences, University of Science and Technology of China, Hefei 230026, China}



\begin{abstract} 
Rotating Radio Transients (RRATs) are neutron stars emitting sporadic radio pulses.
The unique emission of RRATs has been proposed to resemble those of known pulsar types, such as extreme nulling pulsars or pulsars with giant pulses. 
However, the presence of additional radiation beyond these sporadic pulses remains unclear.
Through high-sensitivity observations and extended tracking, we detected the sequential weak emissions in two RRATs with relatively high surface magnetic fields ($B_{\rm s} > 10^{13}$ G): J1846$-$0257 and J1854+0306. These emissions show peak flux densities of 0.15 and 0.41\,mJy, up to 687 and 512 times weaker than our detected RRAT single pulses, respectively.
The weak emissions contribute small fractions ($\sim$ 16\% and 5\%) to the total radio pulse energy releases, contrasting significantly with giant-pulse pulsars where normal pulses dominate. 
Polarization analysis of J1854+0306 suggests that its sporadic RRAT pulses may originate from intermittent enhanced sparking processes due to magnetospheric evolution. 
Our findings indicate that some RRATs may represent a novel class of pulsars, distinct from any previously known subclass. Further observations of sources with similar rotational properties using high-sensitivity instruments could validate the generality of these hidden emissions.
\end{abstract}



\section{Introduction} \label{sec:intro}
Rotating radio transients (RRATs) were first identified in 2006 through their sporadic, repeating radio pulses~\citep{McLaughlin06}. Since their discovery, over 100 such sources have been catalogued (RRATalog), and their dispersion measures (DMs) suggest a Galactic origin. 
By analysing the greatest common divisor of differences in burst arrival times, it is possible to determine the periods of RRATs. Most measured periods are around one second~\citep{McLaughlin06}, a typical timescale for a rotating neutron star. 
Given their observed plausible increasing periods and inferred high brightness temperatures~\citep{McLaughlin06, McLaughlin09}, it is widely believed that RRATs are neutron stars and are related to pulsars.      

The prevailing definition of an RRAT is a pulsar that is more readily detectable in single pulse searches than in periodicity searches~\citep{Keane11, Cui17}. 
However, this definition pertains more to the detectability of the sources rather than their physical emission mechanisms~\citep{Weltevrede06, Keane11}.
Some more physical definitions suggest that RRATs belong to known pulsar subclasses, such as extreme nulling pulsars~\citep{Biggs92, Wang07} but are only `on' for less than a pulse period~\citep{Burke-Spolaor10, Burke-Spolaor13}, or pulsars with giant pulses~\citep{Kuzmin07} where normal emissions are below the sensitivity of current telescopes~\citep{Weltevrede06}.   
Despite these varying definitions and potential observational biases~\citep{Burke-Spolaor10}, it is essential to investigate the intrinsic properties of RRATs.
Understanding why they are more easily discovered through their single pulses, and whether hidden emissions exist beyond their sporadic pulses, is crucial for exploring their unique radiation mechanisms and origins.

To determine if typical RRATs exhibit `on' states for sequential pulse trains, or show `normal' emissions in high-sensitivity observation, we carried out observations using the Five-hundred-meter Aperture Spherical Radio Telescope (FAST)~\citep{Jiang19}, which provides both high-sensitivity and extended tracking capabilities. Previous studies have suggested that RRATs typically have higher magnetic fields than ordinary pulsars, potentially related to their unusual emission behaviour~\citep{McLaughlin09, Enoto19}. Therefore, our study focuses on two RRATs with relatively high surface magnetic field ($B_{\rm s} > 10^{13}$ G), J1846$-$0257 and J1854+0306, both of which can be tracked by FAST for more than one hour. For comparison, we also observed two other RRATs, J1839$-$0141 and J1913+1330, which have characteristic pulsar surface magnetic field ($B_{\rm s}$ $\sim 10^{12}$ G).  

In this paper, we present the detection and analysis of radio pulse emissions from four RRATs using the FAST telescope. The observation details and data analysis are described in Section~\ref{sec:obs}. Section~\ref{sec:results} presents the properties of the detected emissions, while Section~\ref{sec:dis} presents a discussion of our findings.

\section{Observation and data reduction} \label{sec:obs}

Observations were conducted using the 500\,m$-$diameter FAST radio telescope. The observations were on December 21, 2019, for 1 hour each for J1839$-$0141 and J1846$-$0257, on December 25, 2019, for 1 hour for J1854+0306, and on December 16, 2019, for 3 hours for J1913+1330~\footnote{Due to the relatively small field of view and limited sky coverage of the FATS telescope, we selected four sources with well-constrained timing solutions. As we considered the $B_{\rm s}$ as a key point for RRATs study, our sample includes the only two RRATs with relatively high $B_{\rm s}$ (i.e., J1846$-$0257 and J1854+0306) that could be effectively monitored using FAST. The other two RRATs with relatively low $B_{\rm s}$ (i.e., J1839$-$0141 and J1913+1330) were chosen as representative examples, and we plan to propose observations of more similar sources in the future. Given the limited allocation of FAST observing time, the maximum observation length for most sources was around 1 hour, except for J1913+1330, which had up to 3 hours of tracking time from a separate dedicated project~\citep{Zhang23}.}.
Dual linear polarization signals were 8-bit sampled and channelized~\citep{Jiang20} using the Reconfigurable Open Architecture Computing Hardware generation 2 (ROACH 2)~\citep{Hickish16} and stored in PSRFITS search mode format~\citep{Hotan04}. The sample time is 49.153$\mu$s. 
We employed the 19-beam receiver~~\citep{Jiang20} covering a frequency range of 1000$-$1500\,MHz with 4096 channels. 
Signals with four polarizations from the central beam of the multibeam receiver were recorded.
Before each observation, a noise-switched signal was recorded for data calibration.

The FAST search mode data for the four RRATs were folded using the {\sc dspsr} package~\citep{Straten11}, based on the timing ephemeris obtained from the Australia Telescope National Facility (ATNF) pulsar catalogue~\citep{Manchester05}. Initial dedispersion used DM values from the catalogue, which were subsequently refined using high-time-resolution data from bright individual pulses.
The dedispersed polarization data were calibrated using the {\sc psrchive} program~\citep{Hotan04} {\sc pac}, correcting for differential gain and phase between receivers with a noise diode signal injected before each observation.
Channels affected by strong radio frequency interference (RFI) were weighted to zero using the {\sc paz} package~\citep{Hotan04}.

Rotation measures (RMs) for the integrated pulse of each observation were determined using the {\sc rmfit} program~\citep{Hotan04}, searching for a peak in the linearly polarized flux $L = \sqrt{Q^2 + U^2}$ within an RM range from $-4000$ to 4000\,rad\,m$^{2}$, with a step of $1$\,rad\,m$^{2}$. {\sc rmfit} corrected for Faraday rotation for each trial RM, producing a total linear polarization profile and an RM spectrum. A Gaussian fit was applied to determine the optimal RM and its 1$\sigma$ uncertainty.
Polarization data from all channels were then corrected for the best-fitting RMs using the {\sc psrchive} program~\citep{Hotan04} {\sc pam} before integration. The folded polarization and intensity data generated by the {\sc dspsr} package~\citep{Straten11} were saved with bins of 8192, 4096, 2048, 1024, 512 and 256~\footnote{Our pipeline processes datasets with different bins to improve sensitivity to single pulses with variable widths. Currently, we applied six bins based on available computing resources, but using more bins would be encouraged.}. 

A threshold-based search identified single pulse candidates in the time series at different resolutions.
Signals with a peak flux-to-noise ratio greater than five were considered single pulse candidates.  
To investigate the presence of hidden emissions in the observed RRATs, we removed pulse trains with single pulse candidates identified in any of the five different time resolutions.
Compared to previous work~\citep{Zhang23}, relaxed criteria were used to select single pulse candidates, which could potentially include false signals.
However, further analysis focused on examining emissions from pulse stacks excluding all identified single pulse candidates.

Due to the presence of noise, linear polarization $L$ tends to be overestimated. To obtain an unbiased estimate of the linear polarization fraction, we used $L_{\rm unbias}$~\citep{Everett01}:  
\begin{equation}
L_{\rm unbias} =
\begin{cases}
\sigma_{I}\,\sqrt{\frac{L}{\sigma_{I}}-1}  & \text{if } \frac{L}{\sigma_{I}} \geq 1.57 \\
0 & \text{otherwise}
\end{cases}
\end{equation}
where $\sigma_{I}$ represents the off-pulse standard deviation in Stokes I.

The radiometer equation was used to estimate the expected root-mean-square (RMS) of the off-pulse data in Jy~\citep{Lorimer04}:
\begin{equation}
S_{\rm sys}=\frac{T_{\rm sys}}{G \sqrt{{\Delta}{\nu}N_p{t}_{\rm obs}}},
\label{equ:limit}
\end{equation}
where $T_{\rm sys} \sim 20{\rm K}$ is the system temperature, $G \sim 16 {\rm K/Jy}$ is the telescope antenna gain~\citep{Jiang20}, ${\Delta}{\nu} = 500$\,MHz is the observing bandwidth, $N_p =2$ is the sum of polarizations, and $t_{\rm obs}$ is the specified time resolution.
By calibrating the data using $S_{\rm sys}$, we derived flux for individual pulses and integrated pulses. 
Fluence was computed by integrating pulse flux above the baseline.

\section{Results} \label{sec:results}

Using the high sensitivity of FAST, we detected high rates of single pulses from all four RRATs. After excluding regions with detected single pulse, RRAT J1846-0257 and 1854+0306, which have higher $B_{\rm s}$, show sequential weak emissions in addition to the RRAT single pulses. We also analysed the full polarization properties of these newly observed weak emissions.

\subsection{RRAT single pulses}

During observations lasting 1 hour each for J1839$-$0141, J1846$-$0257, and J1854+0306, and 3 hours for J1913+1330, we identified 61, 72, 325, and 717 single pulses, respectively, out of 3657, 803, 789, and 11661 rotating periods (Table~\ref{table:properties}). 
We obtained newly determined values for the DMs and RMs of J1839$-$0141, J1846$-$0257 and J1854+0306.
The pulse rate for J1839$-$0141 was $\sim$ 61 hr$^{-1}$, significantly higher than the discovery observation rate of $\sim$ 0.6 hr$^{-1}$~~\citep{McLaughlin06}. 
It was, however, lower than the peak rate of $\sim$ 116 hr$^{-1}$ observed during a 15-minute session that captured its dense pulse train~~\citep{Zhou23}, which also appeared in our observation (Figure~\ref{fig:stck_regB}).
Similarly, J1846$-$0257 exhibited a burst rate of $\sim$ 72 hr$^{-1}$, much higher than its discovery rate of $\sim$ 1.1 hr$^{-1}$~~\citep{McLaughlin06} and slightly surpassing the previously highest reported rate of $\sim$ 60 hr$^{-1}$~~\citep{Zhou23}.
J1854+0306 had a burst rate of $\sim$ 325 hr$^{-1}$, which is significantly higher than its discovery rate of $\sim$ 84 hr$^{-1}$~~\citep{Deneva09} and exceeds the highest previously reported rate of $\sim $216 hr$^{-1}$~~\citep{Zhou23}.
A detailed analysis of J1913+1330's single pulse properties has been presented in a companion paper~~\citep{Zhang23}. 
In this study, its burst rate was $\sim$ 239 hr$^{-1}$, notably higher than its discovery rate of $\sim$ 4.7 hr$^{-1}$~~\citep{McLaughlin06} and slightly exceeding the previously highest rate of $\sim$164 hr$^{-1}$~~\citep{Zhou23}.  
Due to the increased sensitivity of FAST's 19-beam receiver, our observations detected higher pulse rates for these sources than previously reported observations.
Single pulses occurring sequentially with a waiting time of one rotation period were detected in all four RRATs (Figure~\ref{fig:stck_highB} and \ref{fig:stck_regB}).
As shown in Table~\ref{table:properties}, among the identified single pulses, the minimum flux densities of J1839$-$0141, J1846$-$0257, J1854+0306 and J1913+1330 were 4.18, 2.02, 2.18 and 2.32\,mJy, respectively, while the maximum flux densities were 286.76, 103.14, 210.10 and 1498.38\,mJy, respectively.   

\begin{table*}
\caption{\bf FAST observations of four RRATs.}
\renewcommand\arraystretch{1.2}
\begin{center}
\begin{threeparttable}
\begin{tabular}{c|cccc}
\hline
\hline
{\bf Name} & {\bf J1839$-$0141} & {\bf J1846$-$0257} & {\bf 1854+0306} & {\bf J1913+1330} \\
\hline
T$_{\rm obs}$ (hr) & 1 & 1 & 1 & 3 \\
P (s) & 0.933 & 4.477 & 4.558 & 0.923 \\
B$_{\rm s}$ (10$^{12}$ G) & 2.38 & 27.1 & 26.0 & 2.86 \\
DM (pc\,cm$^{-3}$)& 294.3(1)\tnote{*} &  236.6(2)\tnote{*} & 196.93(2)\tnote{*} & 175.25(2) \\
RM (rad m$^{-2}$) & 355.1(9)\tnote{*} & 214.8(4)\tnote{*} & $-$41.6(6)\tnote{*} & 932.8(6.6) \\
N$_{\rm s}$/N$_{\rm p}$ \tnote{a} & 61/3657 & 72/803 & 325/789 & 717/11661 \\
F$_{\rm s,max}$ (mJy) \tnote{b} &286.76&103.14&210.10&1498.38 \\
F$_{\rm s,min}$ (mJy) \tnote{b} &4.18&2.02&2.18&2.32 \\
F$_{\rm w,mean}$ (mJy) \tnote{b} &$<$0.03&0.15&0.41&$<$0.03 \\
Dist (kpc) \tnote{c} & 6.099/6.279 & 4.001/5.218 & 4.555/5.313 & 6.171/5.705 \\ 
L$_{\rm weak}$ (mJy kpc$^2$) \tnote{c,d}\;\; & $<$13.9/14.7 & 30.2/51.4 & 104.3/141.9 & $<$14.4/12.3 \\
\hline
\end{tabular}
   \begin{tablenotes}
        \footnotesize
        \item[*] Newly determined values in this paper.
        \item[a] The periods affected by strong FRI were removed.
        \item[b] Peak flux density.
        \item[c] The estimation used the YMW16 model \citep{YMW16} or, where indicated with a ``/ '', the NE2001 model \citep{ne2001}.
        \item[d] Isotropic-equivalent spectral luminosity. 
     \end{tablenotes}
\end{threeparttable}
\end{center}
\label{table:properties}
\end{table*}

\subsection{Sequential weak emissions}
\begin{figure*}
\centering
\begin{tabular}{llll}
\includegraphics[height=8.5cm]{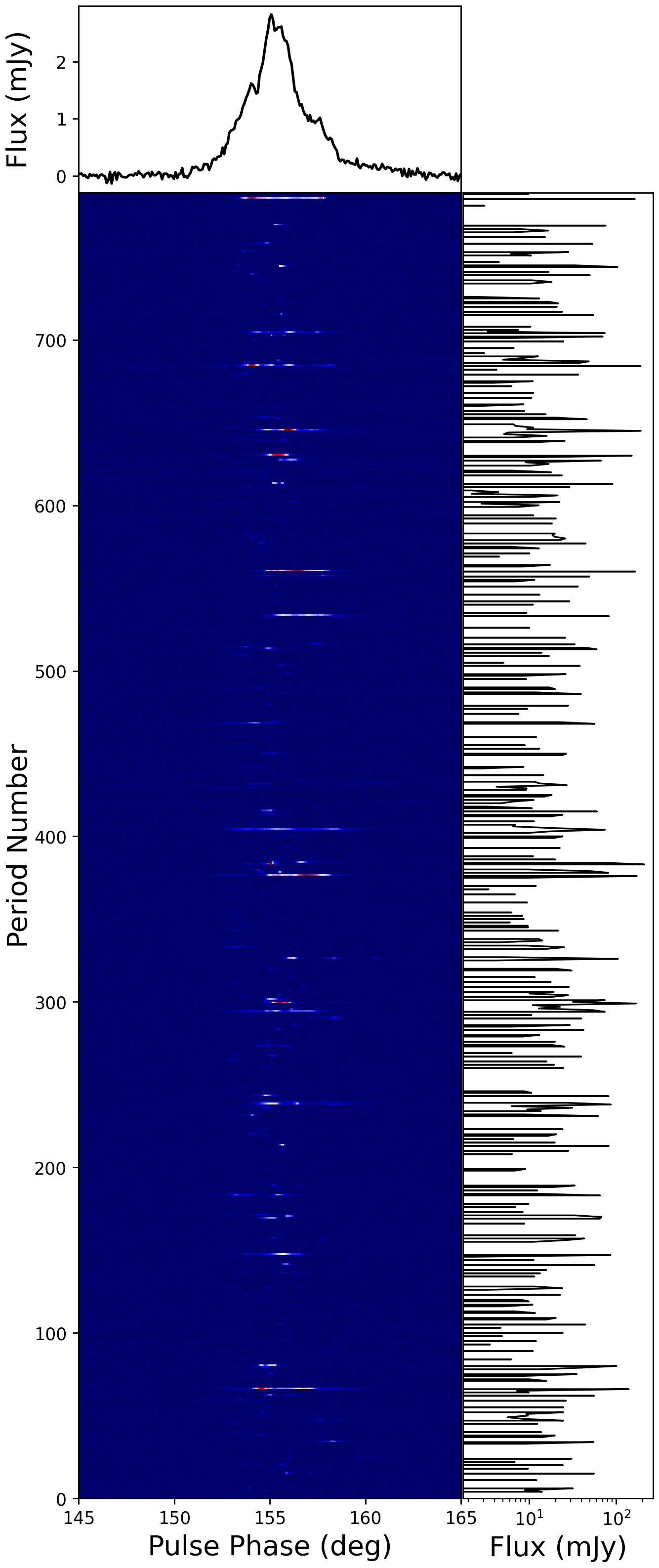} &
\includegraphics[height=8.5cm]{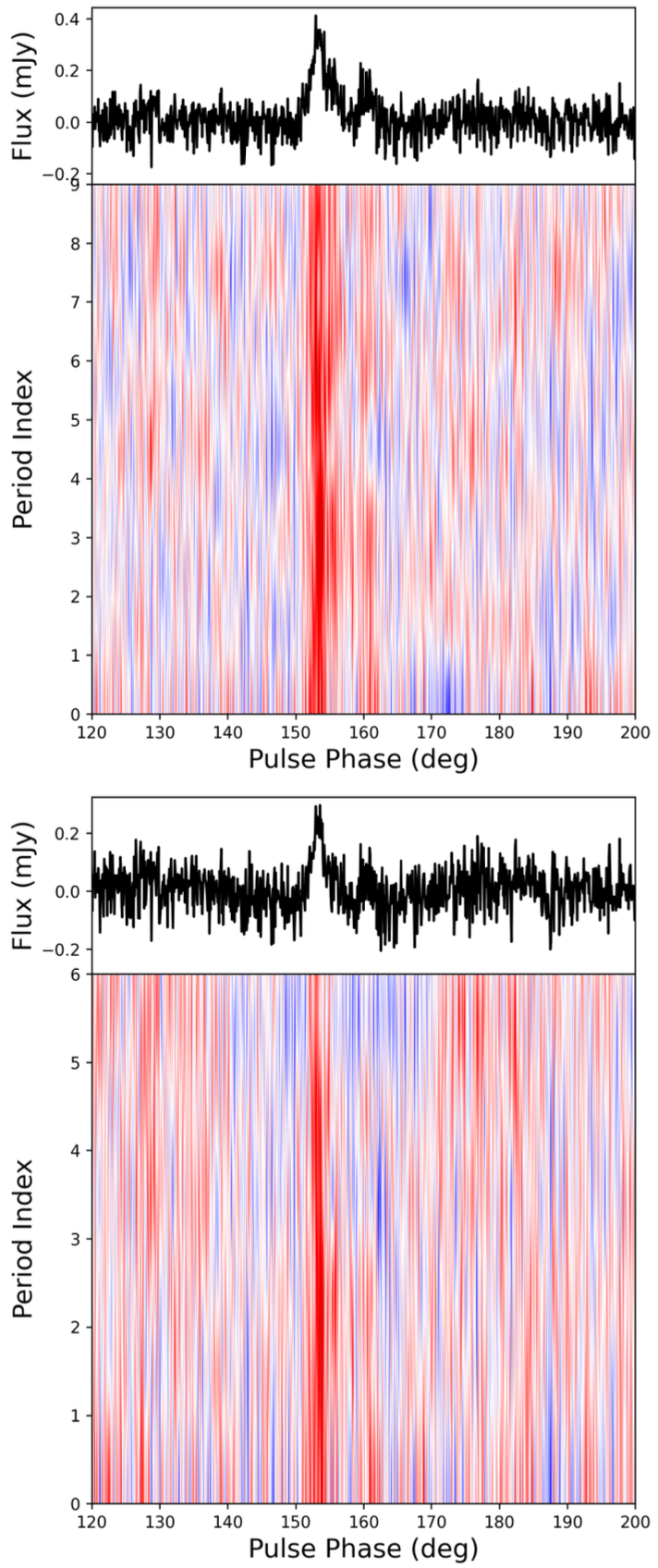} &
\includegraphics[height=8.5cm]{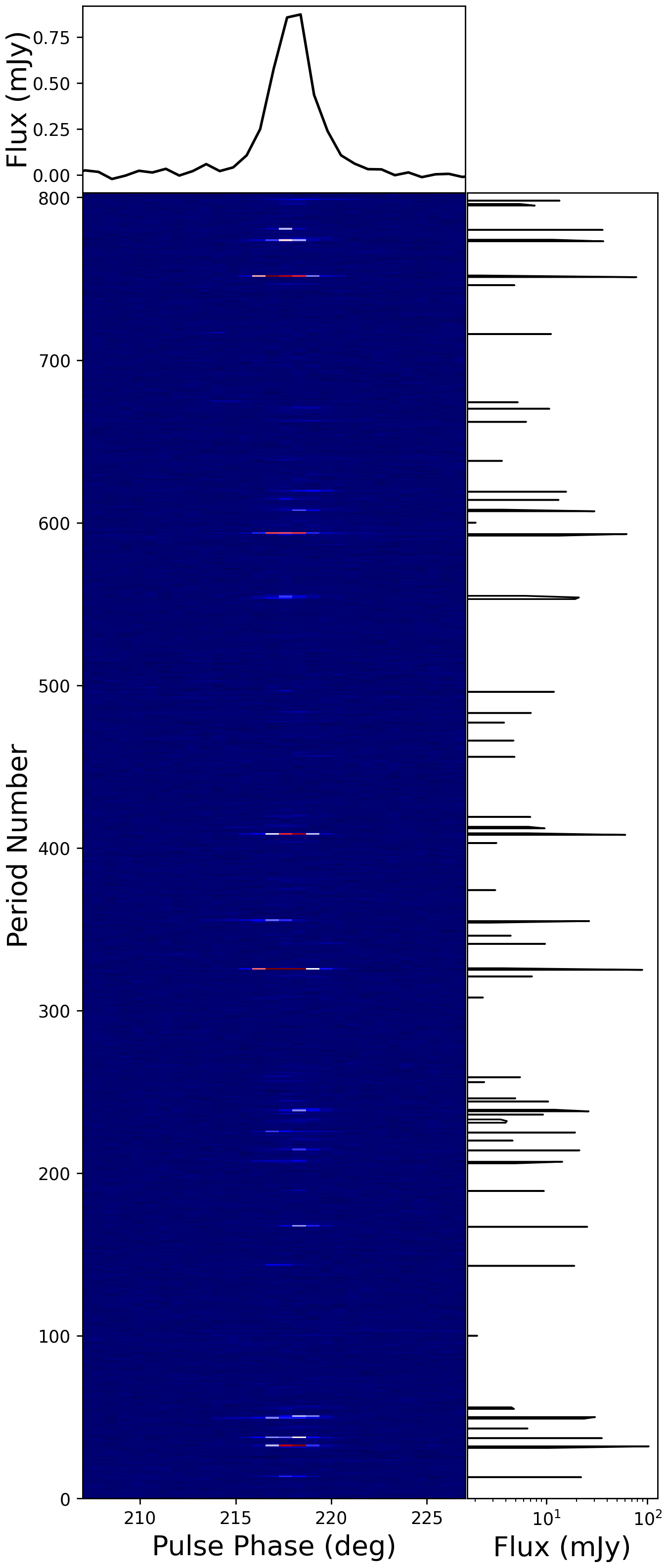} &
\includegraphics[height=8.5cm]{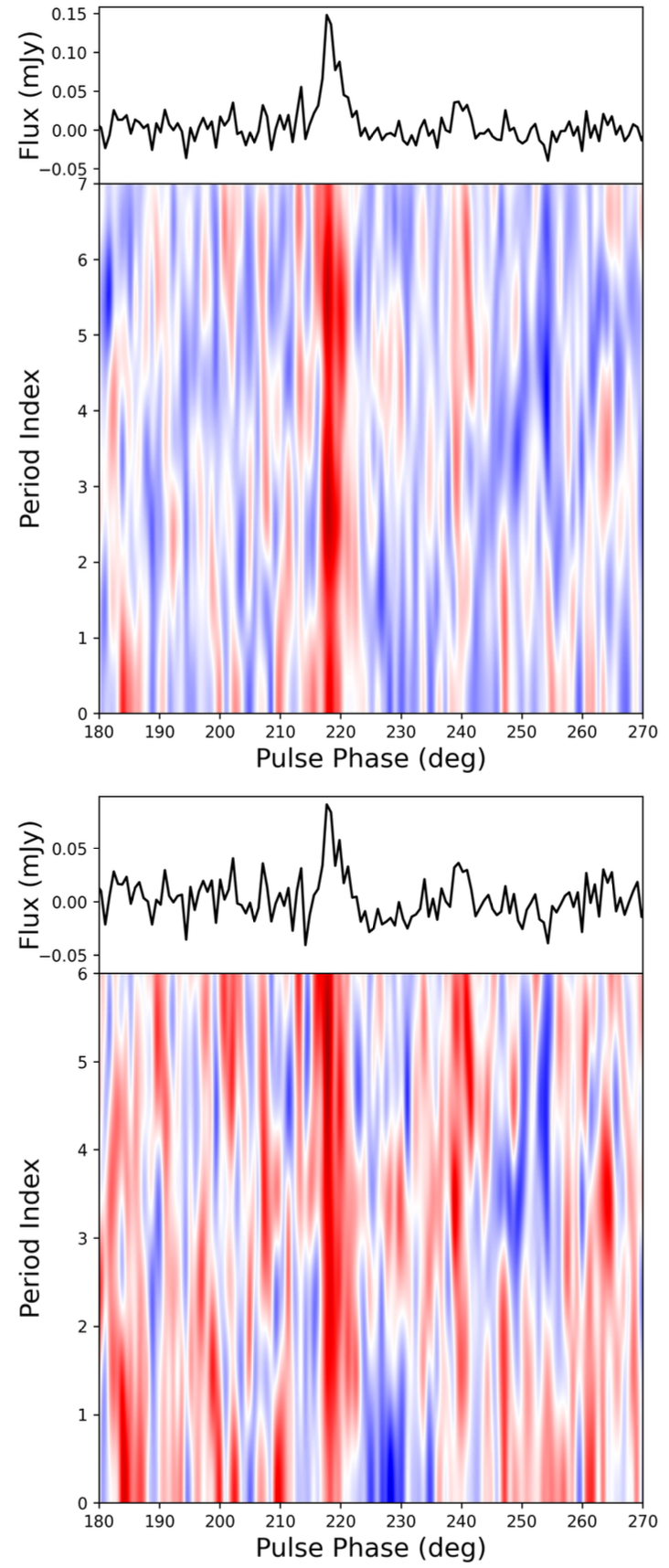} 
\end{tabular}
\caption{\textbf{Pulse trains of J1854+0306 (left) and J1846-0257 (right):} For each panel, left is the train of individual pulses for all periods, with the peak flux of each single pulse plotted immediately to the right; the top-right shows the train of pulses for periods with a peak flux-to-noise ratio $<$ 5; the bottom-right is the train of pulses for periods with a peak flux-to-noise ratio $<$ 3. The stacks of the weak emissions for J1854+0306 and J1846-0257 are folded over 50 and 100 rotation periods, respectively. The Y-axis labelled ``Period Index'' corresponds to the folded time index based on pulsar periods.
Each subplot also presents the mean profile at the top.}
\label{fig:stck_highB}
\end{figure*}

\begin{figure*}
\centering
\begin{tabular}{cc}
\includegraphics[width=5cm]{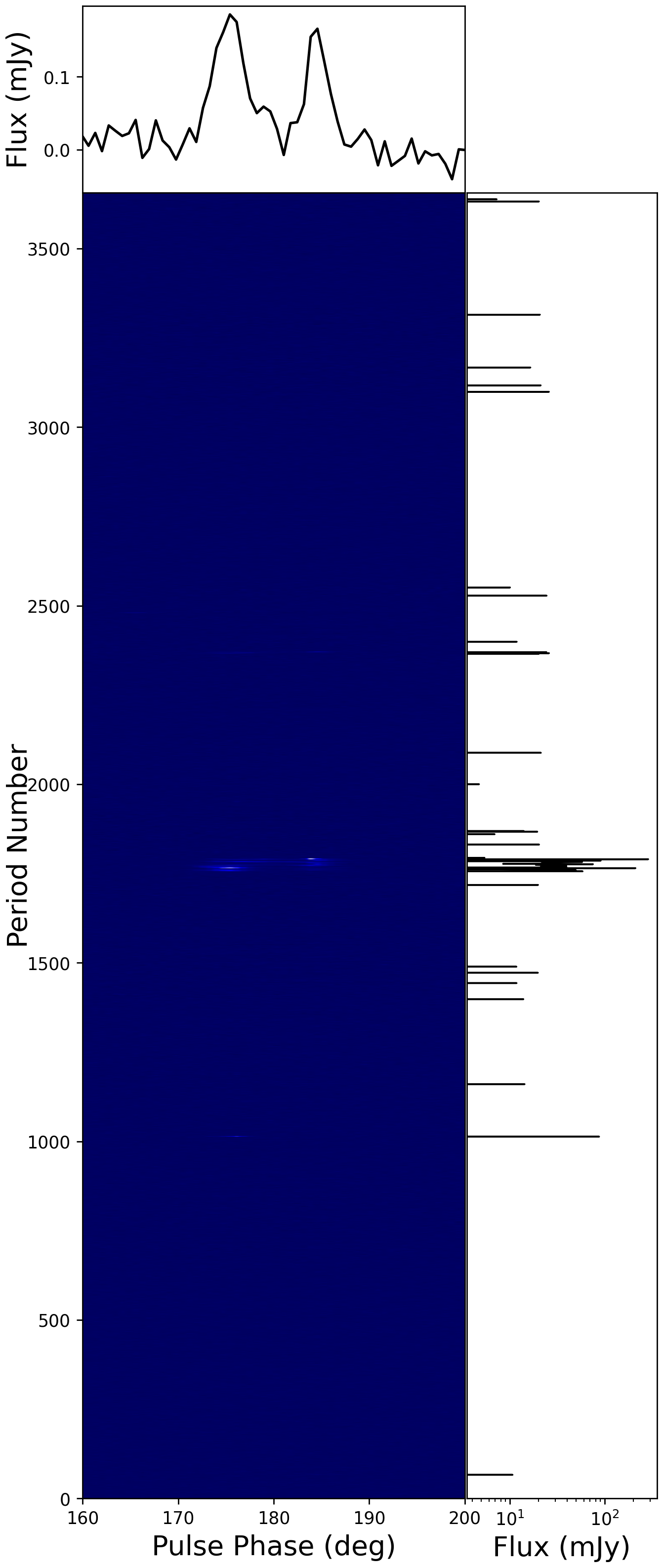} &
\includegraphics[width=5cm]{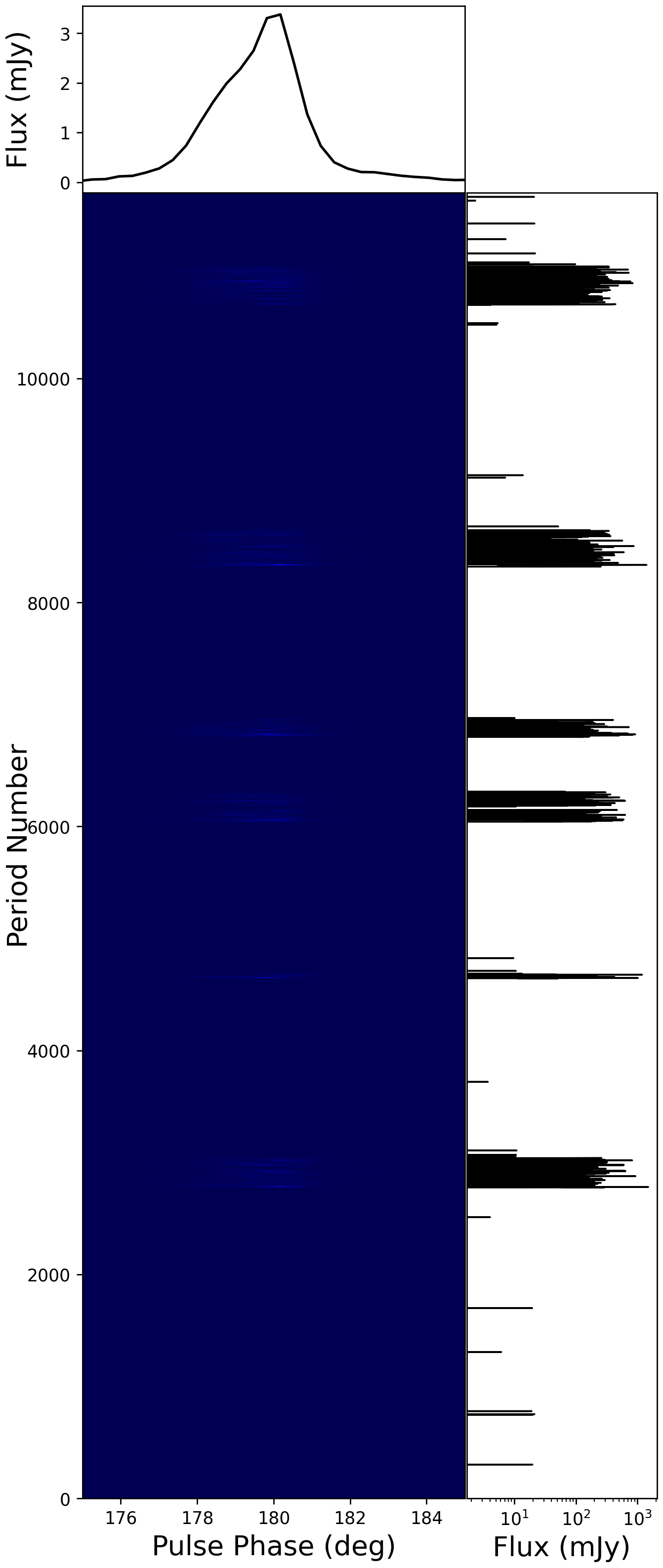} \\
\includegraphics[width=5cm]{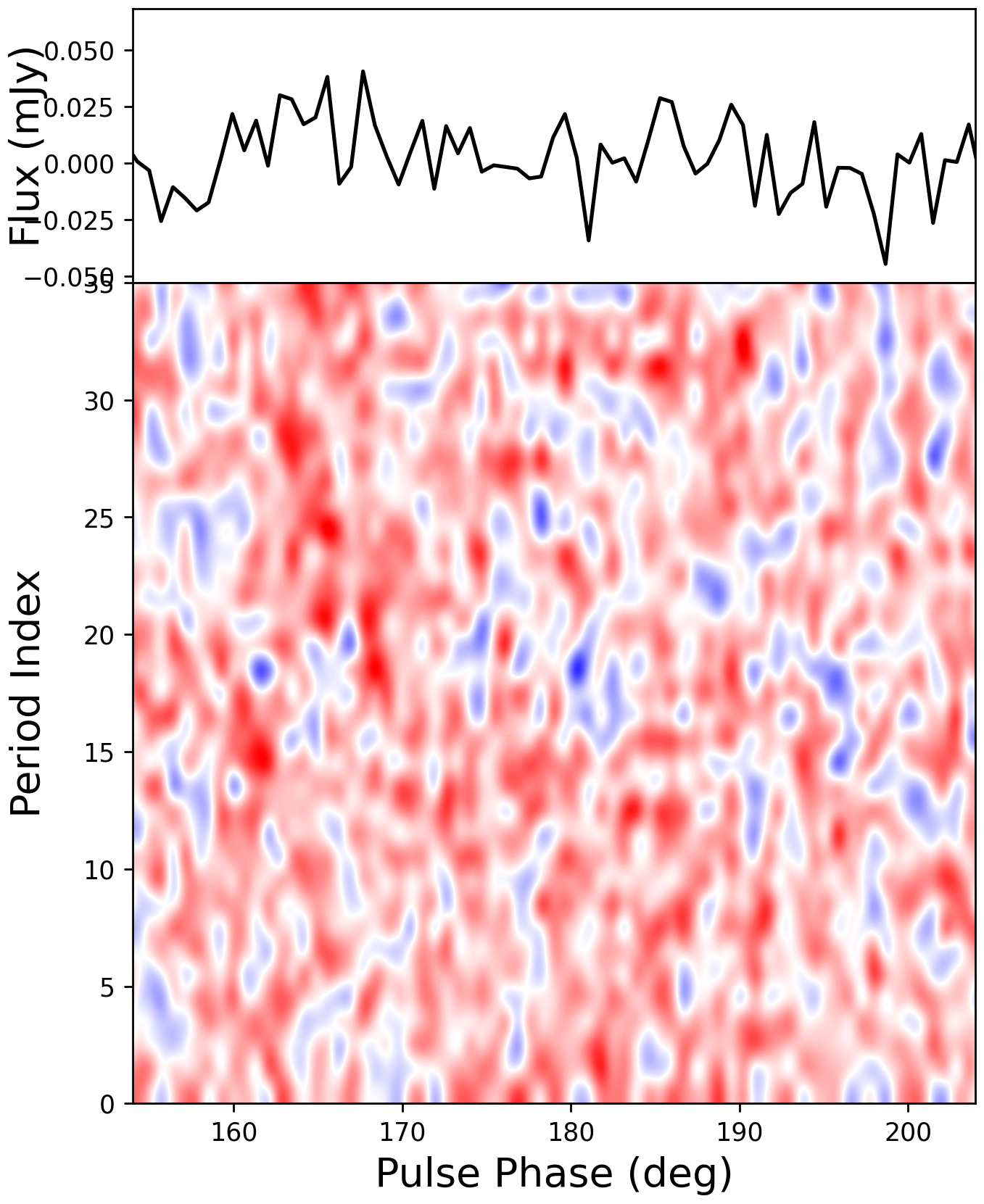} &
\includegraphics[width=5cm]{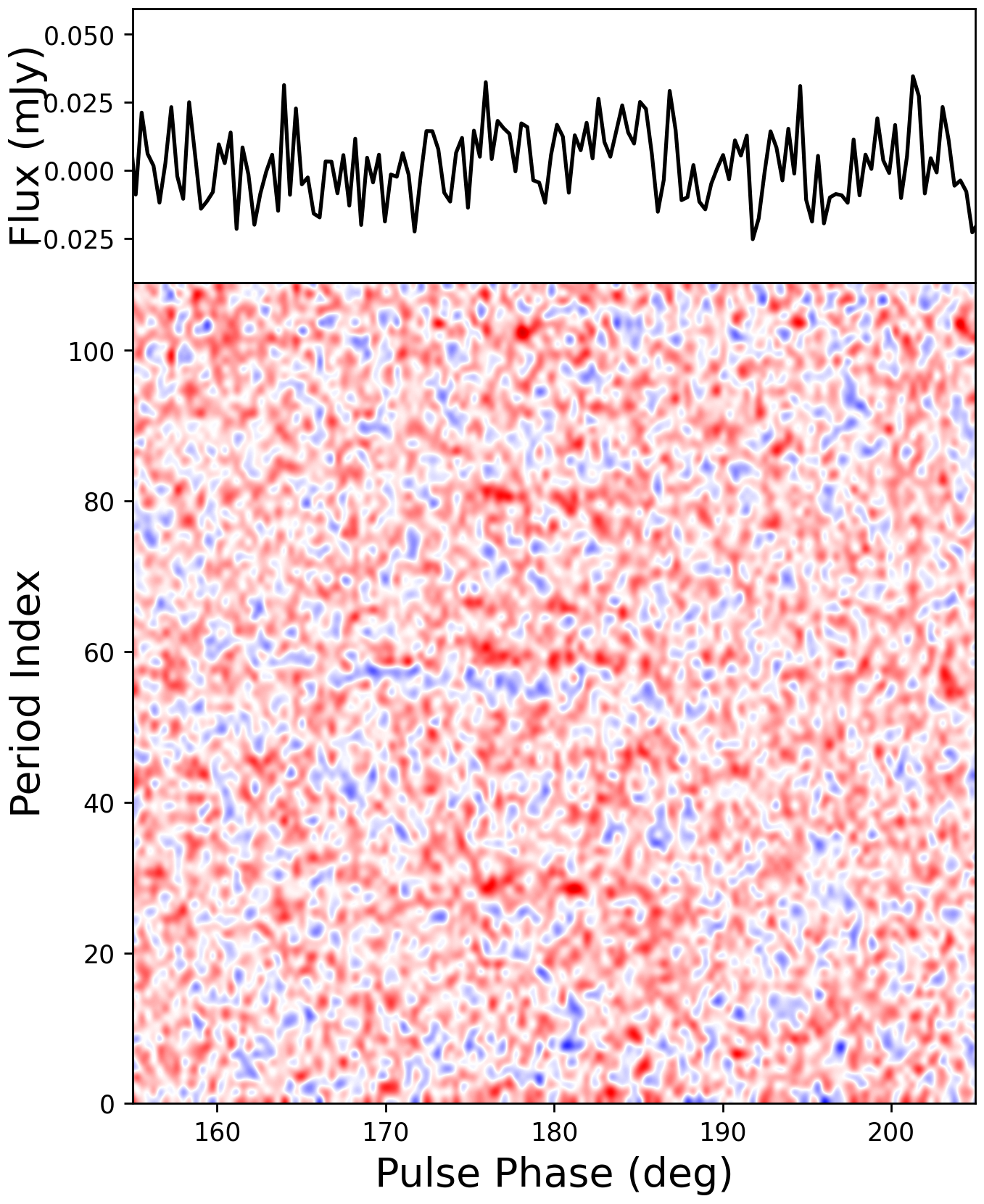} \\
\end{tabular}
\caption{\textbf{Pulse trains of J1839$-$0141 (left) and J1913+1330 (right):} For each panel, the top displays the train of individual pulses for all periods, with the peak flux of each single pulse plotted immediately to the right; the bottom is the train of pulses for periods with peak flux-to-noise ratio $<$ 5. The stacks of the weak emissions for J1839$-$0141 and J1913+1330 have been folded over 100 rotation periods. The Y-axis labelled ``Period Index'' corresponds to the folded time index based on pulsar periods.
Each subplot also presents the mean profile at the top.}
\label{fig:stck_regB}
\end{figure*}

Our high-sensitivity observations revealed relatively high rates of single pulses from the observed RRATs, we further investigated the possible presence of latent emissions.
Initially, we excluded all single pulse stacks within the pulse trains. Interestingly, RRATs J1846$-$0257 and J1854+0306, both of which have higher $B_{\rm s}$, showed sequential weak emissions in addition to the RRATs single pulses (Figure~\ref{fig:stck_highB}), with mean flux densities of 0.15 and 0.41\,mJy, respectively. Integrated pulse profiles and plausible sequential signals were clearly observed throughout both entire observations. 
In contrast, RRATs J1839-0141 and J1913+1330, which have lower $B_{\rm s}$, presented no distinct emissions beyond their RRATs single pulses (Figure~\ref{fig:stck_regB}), with mean flux density constraints of $<$0.03\,mJy for both. 
Even after removing all pulse stacks with peak flux-to-noise ratio $>$ 3 for J1846$-$0257 and J1854+0306, sequential weak emissions remained evident (Figure~\ref{fig:stck_highB}). 
Hereinafter, we refer to the detected single pulse as ``strong RRAT emission'', and the newly observed sequential weak emissions as ``weak emission''. 

\subsection{Polarimetric properties of J1854+0306 and J1846$-$0257} 
\begin{figure*}
  \centering
  \includegraphics[width=10cm]{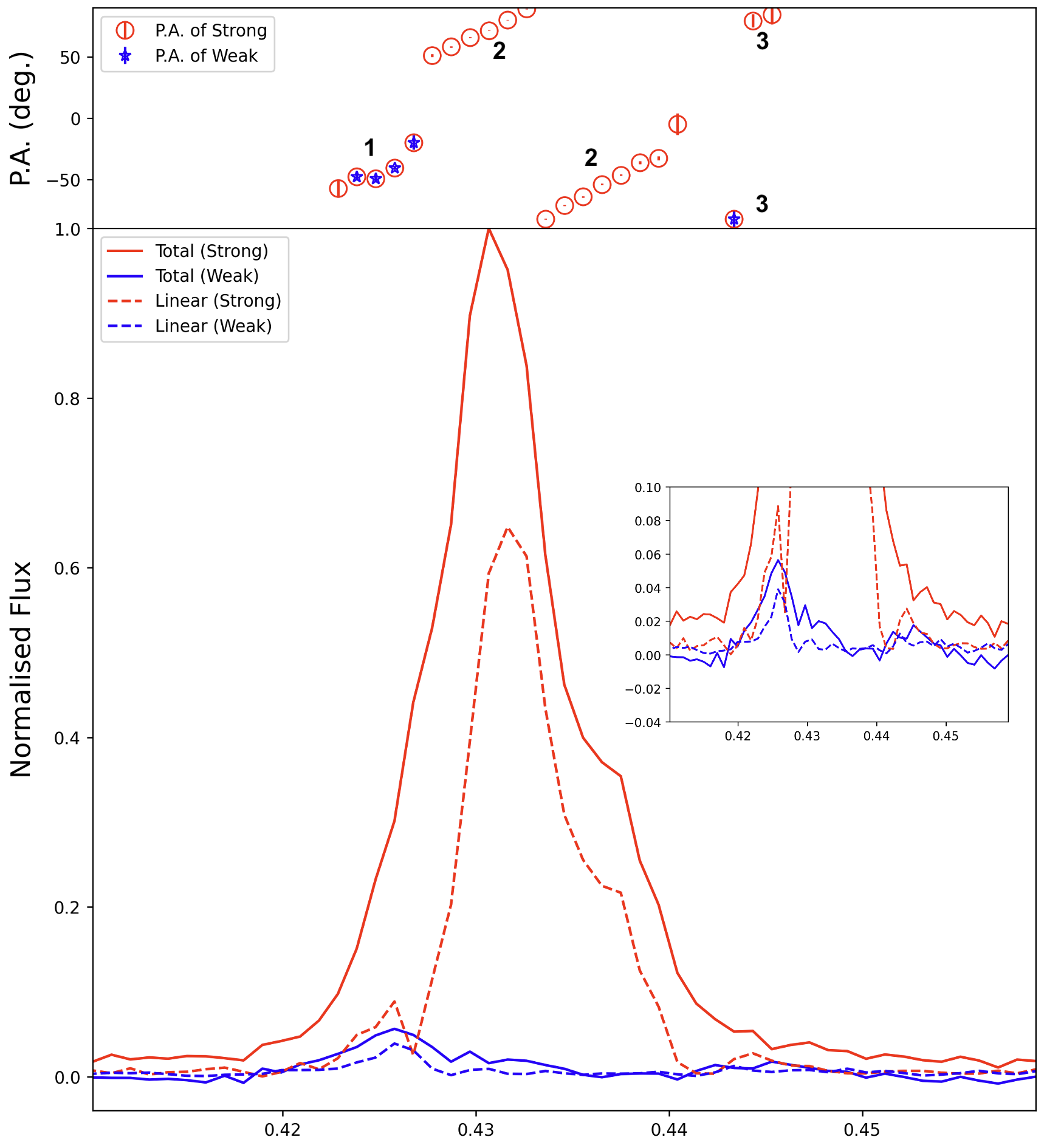}
  \caption{ {\bf Polarization properties of J1854+0306.} The upper panel displays the position angle of linear polarization at 1250\,MHz, with red circles and blue stars representing strong RRAT emission and sequential weak emission, respectively. The component numbers are indicated. The lower panel presents the integrated polarization pulse profile, where the red and blue curves correspond to strong RRAT emission and sequential weak emission. Solid lines represent total intensity, while dashed lines indicate linear polarization. 
  The inset provides a detailed view of the polarization profile of the sequential weak emission.} 
 \label{fig:J1854_pol}
\end{figure*}

\begin{figure*}
\centering
\begin{tabular}{cc}
\includegraphics[height=10cm]{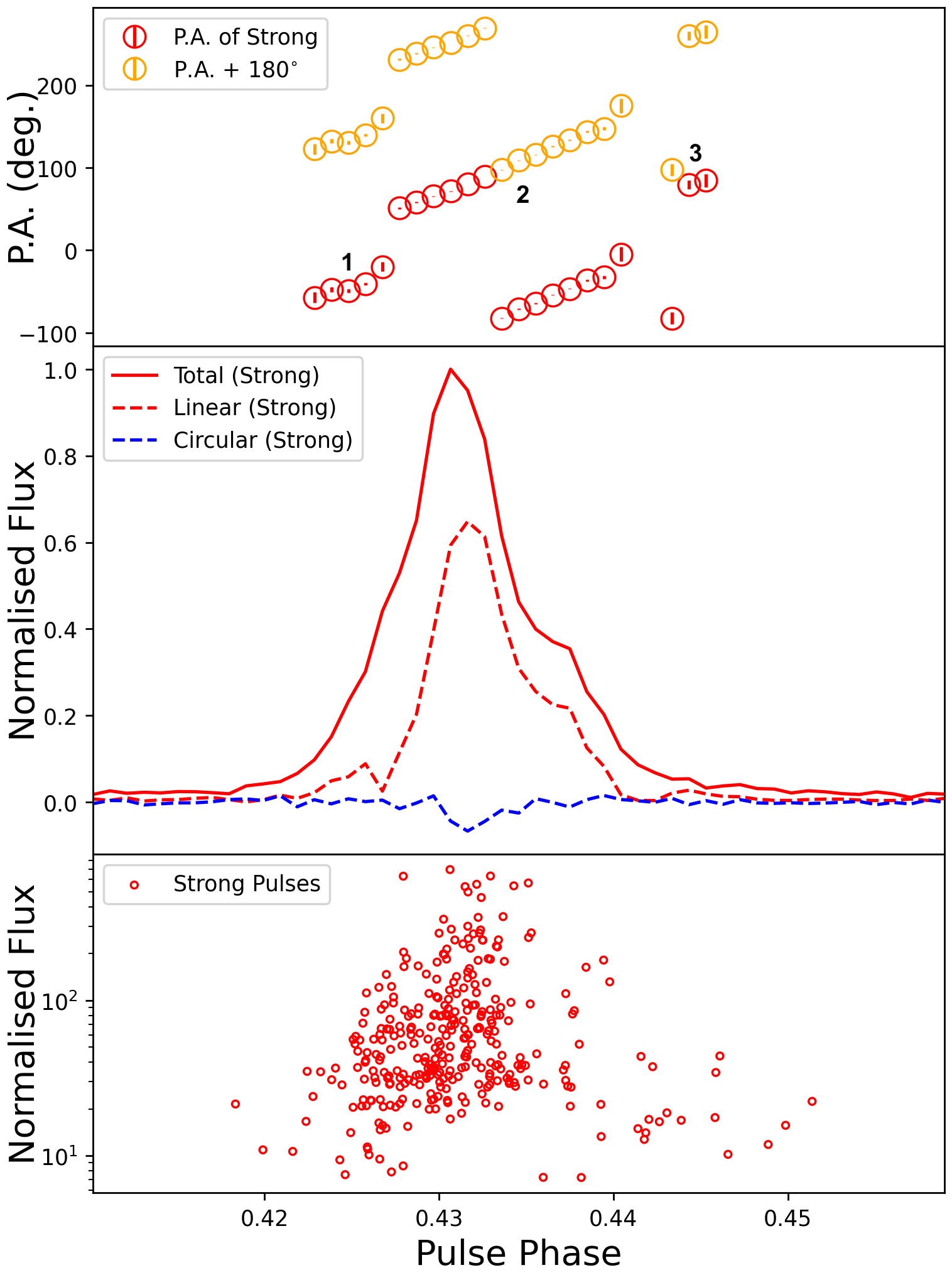} &
\includegraphics[height=10cm]{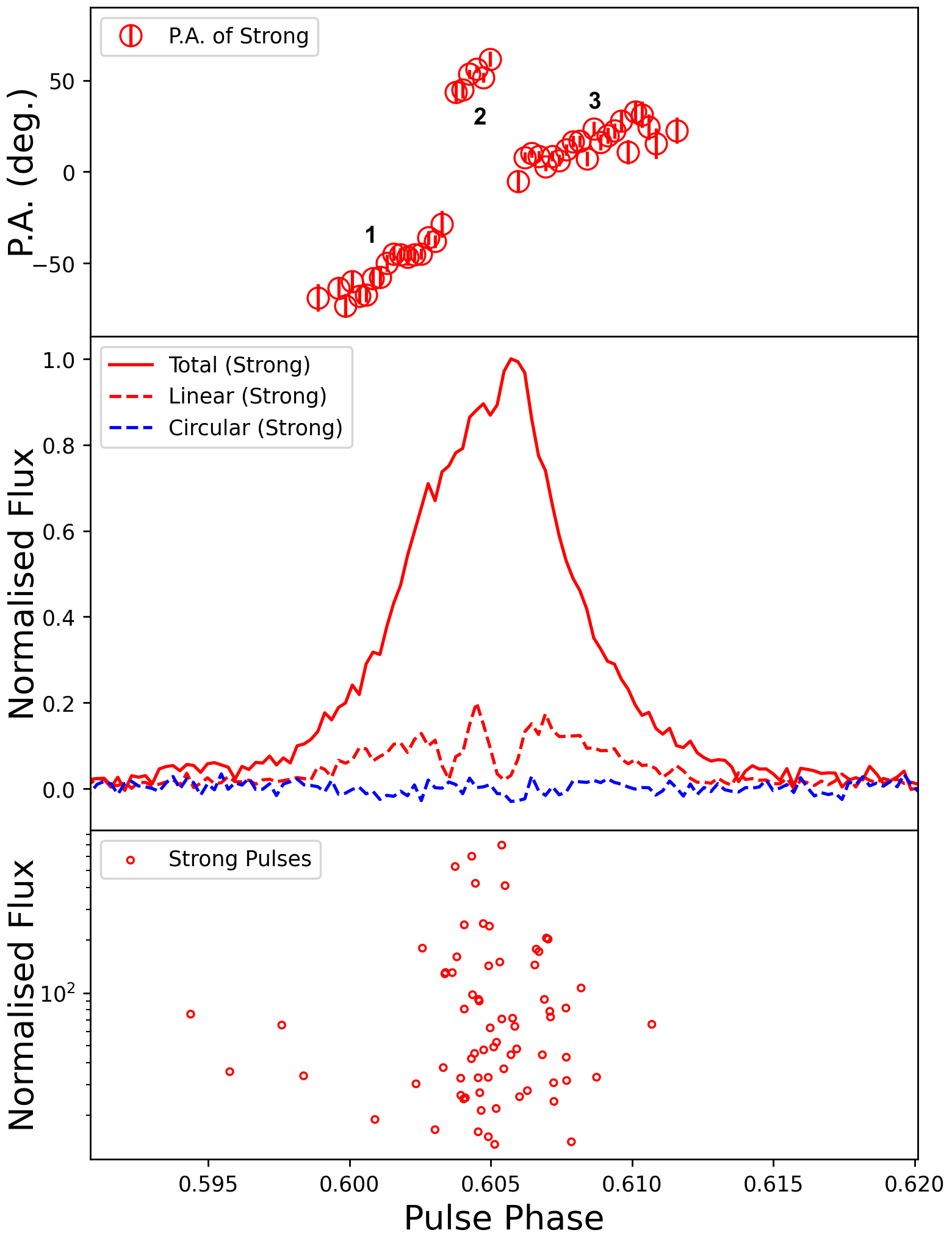} \\
\end{tabular}
\caption{\textbf{Polarization properties of strong RRAT emissions of J1854+0306 and J1846-0257:} In each subplot, the upper panel displays the position angle of linear polarization at 1250\,MHz. The component numbers are indicated. The middle panel presents the polarization pulse profile, with the red solid line, red dashed line and blue dashed line denoting total intensity, linear polarization, and circular polarization. The lower panel is the phase-normalized flux density for individual strong RRAT pulses.
}
\label{fig:pol_strong}
\end{figure*}

After de-dispersing at our best-fitting DM values and correcting the RMs of J1854+0306 and J1846$-$0257, we present the integrated polarized pulse profiles of pulse stacks containing only the strong RRATs emissions in Figure~\ref{fig:pol_strong}.
The linear polarization fractions of the strong RRAT emissions from J1854+0306 and J1846$-$0257 are $45.6 \pm 0.5$\% and $15.8 \pm 0.3$\%, respectively.
The integrated linear polarized pulse profiles and their position angles (PAs) exhibit three distinct components.
The PAs show an abrupt upward jump of $\sim$ 90 degrees between components 1 and 2, followed by a downward jump of a similar magnitude between components 2 and 3. These PA changes differ from the classical rotating vector model (RVM) commonly applied in pulsar emission studies~~\citep{Radhakrishnan69}, but they resemble the special phenomenon of ``orthogonally polarized modes (OPMs)'' observed in some pulsars~~\citep{Manchester75}. 
Further analysis of the weak emissions from these two RRATs revealed detectable linear polarization only for J1854+0306 (Figure~\ref{fig:J1854_pol}). We could not obtain a distinct polarized profile for the weak emissions from J1846$-$0257, likely due to its low flux density and small linear polarization fraction. 
The linear polarization fraction of the weak emission from J1854+0306 is $49.8 \pm 27.6$.
Notably, the integrated linear polarized pulse profile and its PA also show two distinct components, which overlap with the first and third components of the linear polarized pulse profiles and PAs of the strong RRAT emissions.  

\section{Discussion}\label{sec:dis}

\subsection{Emissions beyond sporadic pulses in normal and high$-B_{\rm s}$ RRATs}

\begin{figure*}
\centering
\begin{tabular}{ll}
\includegraphics[height=8cm]{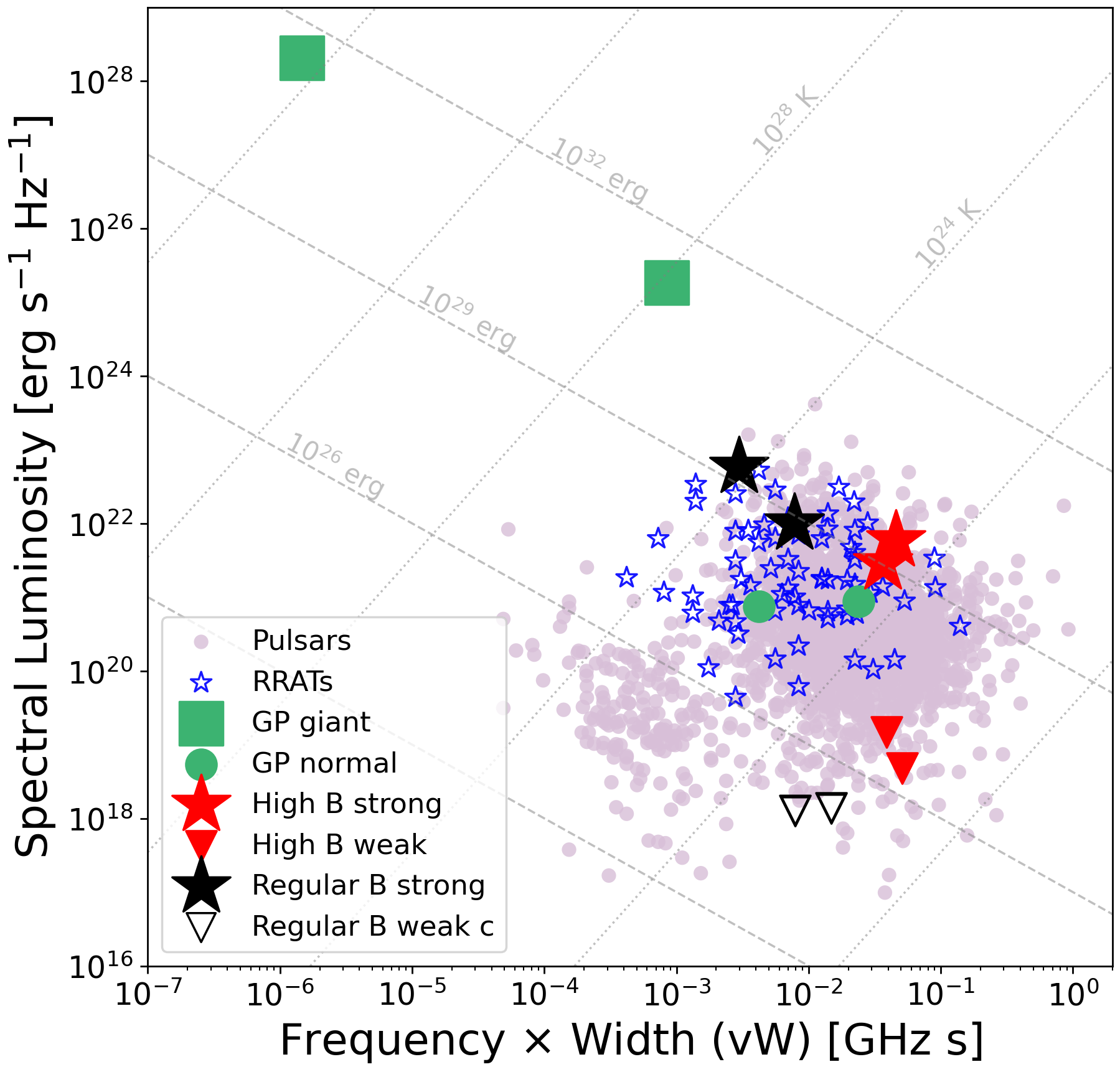}&
\includegraphics[height=6cm]{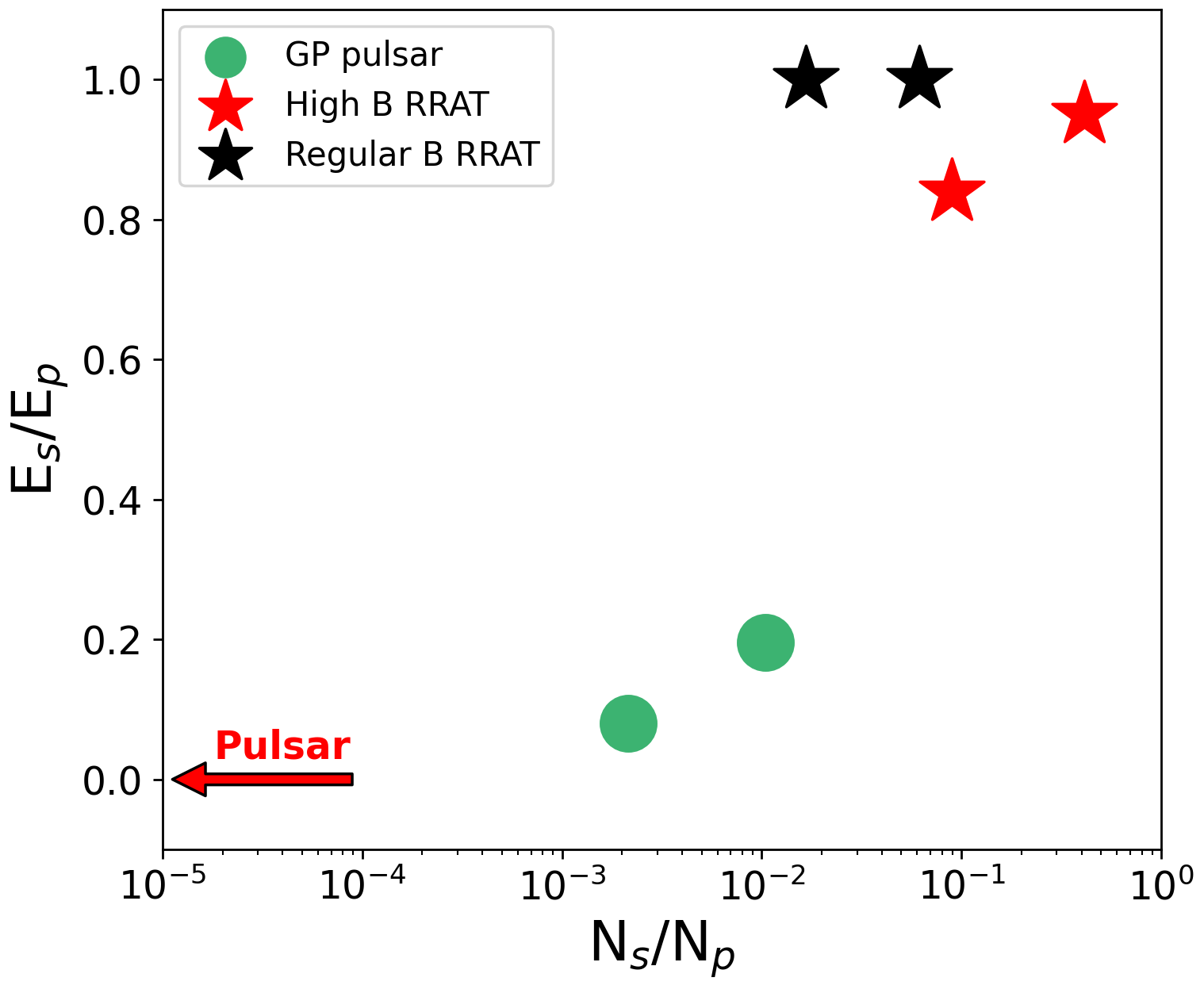}
\end{tabular}
\caption{
    {\bf Left:} Isotropic-equivalent spectral luminosity versus the product of observing frequency and pulse width for various pulsars.
    The brightest strong RRAT pulses from RRATs with relatively high $B_{\rm s}$ (J1846-0257 and 1854+0306), detected in this study, are represented by large red-filled stars, while their weak emissions are denoted by small red-filled triangles.
    The brightest strong RRATs pulses from RRATs with regular $B_{\rm s}$ (J1839$-$0141 and J1913+1330), detected in this study, are shown as large black filled stars, with their weak emissions constraints indicated by small black unfilled triangles.
    The brightest giant pulses burst from J0534+2200~~\citep{Bera19} and J0540$-$6919~~\citep{Geyer21}, are denoted by large green filled squares, with their normal emissions~~\citep{Manchester05} marked by small green filled circles.
    Pulsar~\citep{Manchester05} and RRAT populations (RRATalog) are indicated in thistle and blue, respectively.
    Grey dashed lines represent constant isotropic-equivalent energy release, and grey dotted lines indicate constant brightness temperature.
    {\bf Right:} The energy release contribution fraction of strong pulse emissions versus the rate of strong pulse emissions occurring in one rotation period. RRATs with relatively high $B_{\rm s}$ (J1846$-$0257 and 1854+0306) are shown as red filled stars, while RRATs with regular $B_{\rm s}$ (J1839$-$0141 and J1913+1330) are depicted as black filled stars. Pulsars with giant pulses (J0534+2200~\citep{Popov06} and J0540$-$6919~\citep{Geyer21}) are represented by green filled circles.}
\label{fig:radio_sources}
\end{figure*}

\begin{figure*}
  \centering
  \includegraphics[width=14cm]{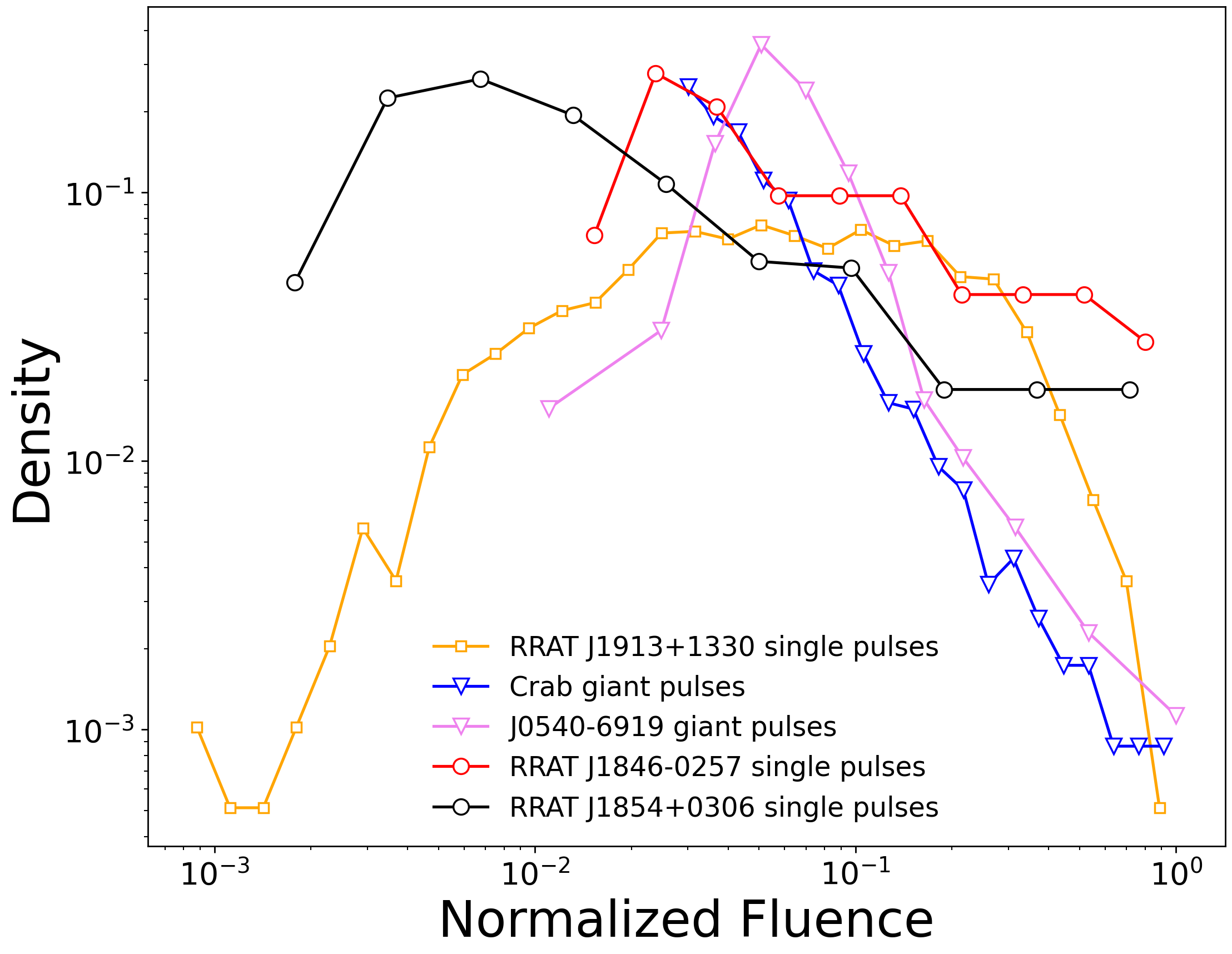}
  \caption{Normalized fluence distributions of single pulses from RRAT J1913+1330, J1846$-$0257 and J1854+0306, as well as giant pulses from the Crab pulsar and J0540$-$6919. The fluence values for each source are normalized to their brightest bursts. Density represents the number of bursts per bin divided by the total detected bursts. } 
 \label{fig:fluence_dis}
\end{figure*}

Initial observations of RRATs did not reveal periodic emissions~\citep{Keane11, Cui17}. During intervals between sporadic pulses, RRAT sources either emit weakly~\citep{Weltevrede06} or not at all (i.e. nulling)~\citep{Burke-Spolaor10, Burke-Spolaor13}. 
The results from our high-sensitivity observations using the FAST telescope support both hypotheses: some RRATs, such as J1839$-$0141 and J1913+1330, appear to lack emissions between detected pulses and have characteristic pulsar surface magnetic fields ($B_{\rm s}$ $\sim 10^{12}$ G). In contrast, other RRATs, like J1846$-$0257 and J1854+0306, exhibit sequential weak emissions in addition to the sporadic pulses and have relatively high $B_{\rm s}$ ($> 10^{13}$ G). 
The estimated isotropic-equivalent spectral luminosities of J1846$-$0257 and J1854+0306 are significantly higher than the luminosity constraints of J1839$-$0141 and J1913+1330 (Table~\ref{table:properties}), suggesting that detecting weak emissions is easier for RRATs with high $B_{\rm s}$ compared to those with regular $B_{\rm s}$.

Notably, \citet{Bhattacharyya18} claimed a “weak persistent emission mode” for J1913+1330 based on relatively low-sensitivity observations. However, \citet{Zhang23} shows that no two distinct emission modes exist for this source based on high-sensitivity FAST observations. Instead, all detected pulses from J1913+1330 seem to originate from a same emission mechanism, exhibiting extreme pulse-to-pulse modulation.  
If the lower-energy part of the single-pulse distribution of J1913+1330 were interpreted as a ``weak persistent emission mode'', its mean flux density would  yield an isotropic-equivalent spectral luminosity of $\sim 10^4$ mJy kpc$^2$, which is about two orders of magnitude higher than that of the ``weak emission'' from the two high $B_{\rm s}$ RRATs. 
Furthermore, in the relatively low-sensitivity observations~\citep{Bhattacharyya18}, the claimed so-called “weak persistent emission mode” of J1913+1330 had variable durations ranging from 2 to 14 minutes and occupied less than 10\% of the observation time. In contract, the observed ``sequential weak emission'' of the two high $B_{\rm s}$ RRATs appears uninterrupted throughout the entire observation (Figure~\ref{fig:stck_highB}).
If the emissions from the two high $B_{\rm s}$ RRATs were similar to those of J1913+1330~\citep{Zhang23}, their energy distributions at high energies, where individual pulses are detectable, should also be comparable. However, as shown in Figure~\ref{fig:fluence_dis}, the fitted indices ($\alpha$) of the power-law-like tail at high energies for RRATs J1846$-$0257 ($\alpha \sim -0.63$) and J1854+0306 ($\alpha \sim -0.64$) are significantly flatter than that of J1913+1330 ($\alpha \sim -2.31$).

\subsection{Comparing newly detected emissions with giant pulse pulsars}

The newly detected weak emissions of J1846$-$0257 and J1854+0306 are much fainter than their strong RRAT emissions, which are up to $\sim$ 688 and $\sim$ 512 times brighter, respectively. This characteristic makes the strong RRAT emissions resemble the so-called ``giant pulses'' observed in some pulsars~\citep{Kuzmin07}.  
For comparison, we examined two pulsars, J0534+2200 (the Crab pulsar) and J0540$-$6919, which exhibit giant pulses with the largest excesses relative to their normal emissions~\citep{Wang19}. As shown in Figure~\ref{fig:radio_sources} (left panel), the normal and strong emissions of these pulsars are significantly higher than those of RRATs with relatively high $B_{\rm s}$.
Furthermore, as shown in Figure~\ref{fig:radio_sources} (right panel), the weak emissions of RRATs J1846$-$0257 and J1854+0306 contribute only small fractions ($\sim$ 16\% and 5\%) to the total radio pulse energy releases. This is a notable difference between the two pulsars with giant pulses, where normal pulses dominate. 
This distinction helps explain why RRATs are more readily discovered through their single pulses rather than through period searches in the Fourier domain.

Furthermore, in Figure~\ref{fig:fluence_dis}, we compare the normalized fluence distributions of single pulses observed from RRAT~J1846$-$0257 and J1854+0306, alongside giant pulses from the Crab pulsar~\citep{Bera19} and J0540$-$6919~\citep{Geyer21}.
Although all sources exhibit a power-law-like tail at high energies, the fitted power-law indices of RRATs J1846$-$0257 ($\alpha \sim -0.63$) and J1854+0306 ($\alpha \sim -0.64$) are significantly flatter than those of the Crab pulsar ($\alpha \sim -1.81$), J0540$-$6919 ($\alpha \sim -2.08$) and RRAT J1913+1330 ($\alpha \sim -2.31$).
This indicates that the energy distributions of the two RRATs with relatively high $B_{\rm s}$ decline much more slowly at high energies compared to the giant pulse pulsars, which may partly explain why their ``strong RRAT emission'' dominates the total radio pulse energy release.

\subsection{Sparking pulsars and the emission mechanisms}

The position angle (PA) swing of linear polarization across the pulse longitude of J1854+0306 supports the magnetospheric origin for the radio emissions~\citep{Radhakrishnan69}. The PA exhibits two orthogonal jumps accompanied by a significant reduction in linear polarization of the strong emission. This behaviour can be understood as a transition between two orthogonally polarized modes as exhibited in other pulsars~\citep{McKinnon98}.

Understanding the temporal behaviors of RRATs is still challenging in the literature~\citep{Keane11,Abhishek22}. The sequential weak emissions in two RRATs with relatively high surface magnetic fields here make it even more puzzling. 
The radio emissions of normal pulsars could be caused by curvature radiation (CR) of relativistic plasma clouds accelerated by the gap above the polar cap region, also called sparking~\citep{Ruderman75}. The characteristics like nulling and pulse drifting in RRATs indicate that the sparking over the polar cap may evolve drastically rather than quasi-steady, potentially induced by the non-uniformity of the crust properties~\citep{Gil00} or the evolving topology of surrounding magnetic fields~\citep{Timokhin07}. Inspired by the partially screened gap (PSG) model aiming to interpret the intriguing radio emissions from magnetars~\citep{Gil07,Medin07,Szary15a,Szary15b}, we suggest that the individual strong pulse and the consecutive emission of the high-$B_{\rm s}$ RRATs could be attributed to distinct sparking regions in different screened states. In the PSG model, outflow of thermal ions from the surface screens the gap by a screening factor of $\eta = 1 - \rho_{\rm i}/\rho_{\rm GJ}$, where $\rho_{\rm i}$ is the charge density of the heavy ions in the gap and $\rho_{\rm GJ}$ is the so-called Goldreich-Julian density~\citep{Goldreich69}. The potential drop of the acceleration gap of the region in the PSG-on mode is thus
\begin{equation}
\Delta V_{\rm on} = \eta \Delta V_{\rm off},  
\end{equation}
where $V_{\rm off}$ is the potential drop of gap in the PSG-off mode.
For regions in the PSG-off mode, the acceleration gap resembles to that of the traditional vacuum gap model~\citep{Ruderman75}, its typical gap height is $\sim 10^4$~cm and it breaks down by the CR of particles. While for sparking regions in the PSG-on mode, its gap height is much smaller ($\sim 10^3$~cm) and the gap breaks down by the Inverse Compton Scattering (ICS) between the particles and thermal photons from the surface~\citep{Szary15a}.

As the RRATs with $B_{\rm s} \sim 10^{13}$~G is in a transition state between the normal pulsar and the magnetars, their polar cap region may consist of patches in both modes. For patches where the surface heating condition is met~\citep{Medin07}, they are PSG-on, otherwise PSG-off. We suppose that the occasionally PSG-on regions produce strong individual pulses and the PSG-off regions produce persistent weak emissions.
Assuming that the coherent curvature emission flux is proportional to the multiplicity $\mathcal{M}$ of the plasma cloud leaves from the gap, we get
\begin{equation}
\frac{F_{\rm strong}}{F_{\rm weak}}  \simeq \eta \frac{\mathcal{M}_{\rm on}}{\mathcal{M}_{\rm off}} \simeq \eta \frac{\langle \gamma_{\rm off}\rangle}{\langle \gamma_{\rm on} \rangle},
\end{equation}
where $\langle \gamma_{\rm on}\rangle$ and $\langle \gamma_{\rm off}\rangle$ represent the average Lorentz factor of the secondary plasma in relevant modes. For typical parameter values of the high-$B_{\rm s}$ pulsar, simulations give that $\langle \gamma_{\rm on}\rangle \in [10^2, 10^3]$ and $\langle \gamma_{\rm off}\rangle \in [10^5, 10^6]$~\citep{Szary13}, making this scenario consistent with the observed flux ratio ($\sim$100) for J1846$-$0257 and J1854$+$0306 shown in Equation (4).

%

\section{Conclusion}

\begin{figure*}
\centering
\begin{tabular}{l}
\includegraphics[height=12cm]{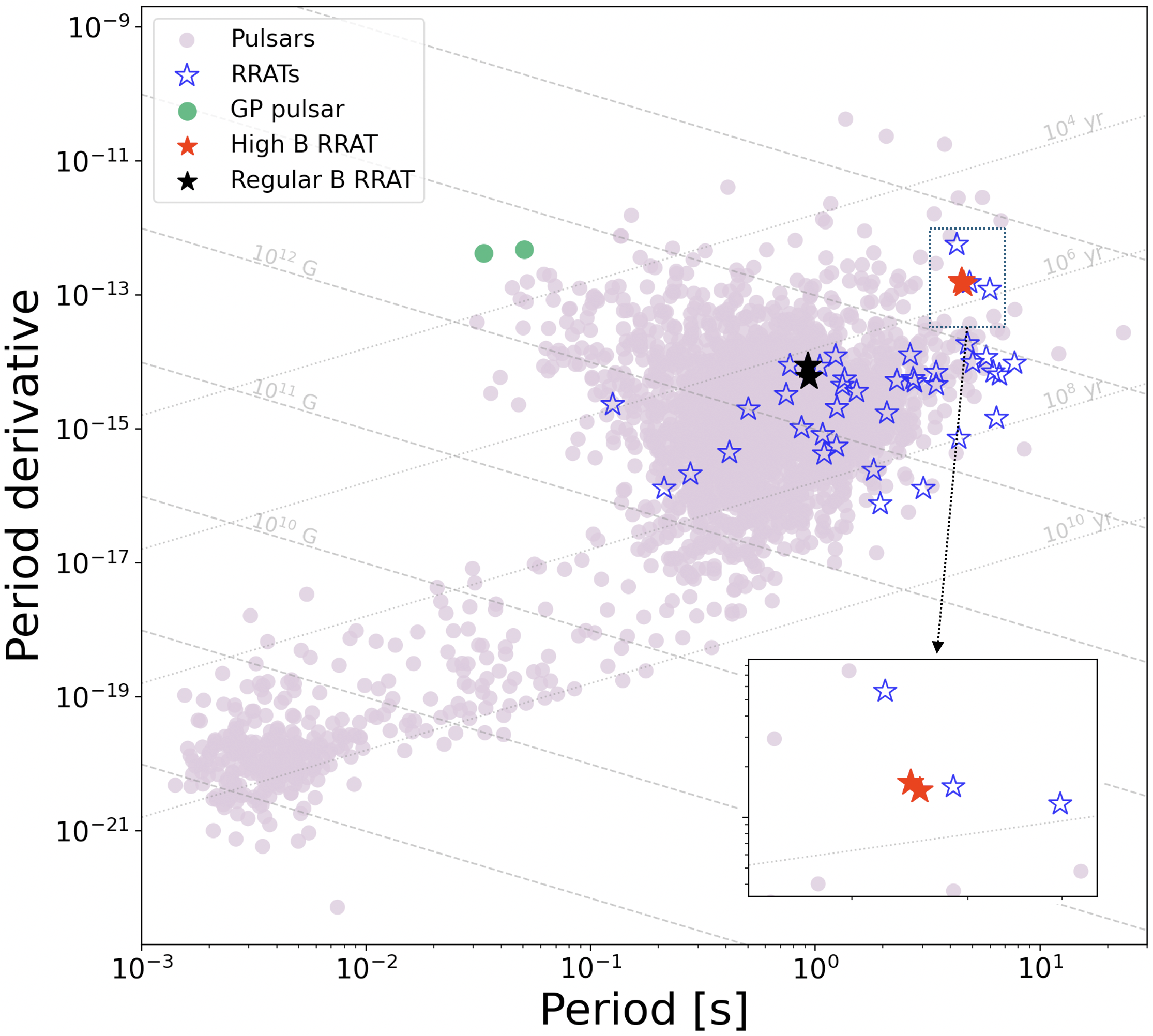}
\end{tabular}
\caption{{\bf Period versus period derivative for pulsars with radio emission, RRATs, pulsars with giant pulses and the RRATs observed in this study.} RRATs with relatively high $B_{\rm s}$ studied in our study (J1846$-$0257 and J1854+0306) are represented by red filled stars, while RRATs with regular $B_{\rm s}$ (J1839$-$0141 and J1913+1330) are denoted by black filled stars. Pulsars with giant pulses (J0534+2200~~\citep{Popov06} and J0540$-$6919~~\citep{Geyer21}) are indicated by green filled circles. 
Grey dashed lines represent constant surface magnetic flux density, while grey dotted lines represent constant spin-down age. The inset gives a closer look at the sources around the RRATs with relatively high $B_{\rm s}$ (J1846$-$0257 and J1854+0306). } 
\label{fig:p0p1}
\end{figure*}

Among the RRATs studied, some have been identified as regular pulsars in follow-up observations~\citep{Burke-Spolaor10}, while others, like J1913+1330, could be similar to nulling pulsars but exhibit extremely variable properties~\citep{Zhang23}. 
Our FAST observations of four RRATs indicate that J1846$-$0257 and J1854+0306, which have relatively high $B_{\rm s}$, display sequential weak emissions during intervals between their sporadic pulses. The estimated luminosities of these emissions are significantly higher than the luminosity constraints from the non-detection of weak emissions in J1839$-$0141 and J1913+1330, which have characteristic pulsar $B_{\rm s}$. 
We suggest that the emission properties of RRATs, at least for J1846$-$0257 and J1854+0306, are distinct from those of normal pulsars and two well-known pulsar subclasses (i.e. nulling pulsars and giant pulse pulsars). Further detailed investigation and classifications are needed to understand this phenomenon.

Our results indicate that detecting weak emissions may be easier for RRATs with high $B_{\rm s}$ compared to those with regular $B_{\rm s}$. Thus, focusing on RRATs with relatively high $B_{\rm s}$ is likely an effective observational strategy.
Unfortunately, out of around 40 RRATs with measured periods and period derivatives for estimating surface magnetic fields, only five have relatively high $B_{\rm s}$ ($> 10^{13}$ G), and among these, only the two RRATs observed in this study could be monitored using the FAST telescope. Other RRATs with similar $B_{\rm s}$, such as J0736$-$6304 and J0847$-$4316 (Figure~\ref{fig:p0p1}), are not currently observable with instruments providing both high sensitivity and extended tracking.   
If RRATs represent a particular stage in pulsar evolution~~\citep{Keane11}, further observations of sources with similar rotational properties using high-sensitivity instruments could help verify the generality of these hidden emissions.

\clearpage
\section*{Acknowledgments}
This work is partially supported by the National SKA Program of China (2022SKA0130100,2020SKA0120300), the National Natural Science Foundation of China (grant Nos. 12041306, 12273113, 12233002, 12003028, 12321003), the international Partnership Program of Chinese Academy of Sciences for Grand Challenges (114332KYSB20210018), the National Key R\&D Program of China (2021YFA0718500), the ACAMAR Postdoctoral Fellow, China Postdoctoral Science Foundation (grant No. 2020M681758), and the Natural Science Foundation of Jiangsu Province (grant Nos. BK20210998). JJG acknowledges the support from the Youth Innovation Promotion Association (2023331).  YPY acknowledges the support from the ``Science \& Technology Champion Project'' (202005AB160002) and from two ``Team Projects'' – the ``Innovation Team'' (202105AE160021) and the ``Top Team'' (202305AT350002), all funded by the ``Yunnan Revitalization Talent Support Program''. \\


\clearpage
\bibliography{biblio}{}

\begin{thebibliography}{}
\expandafter\ifx\csname natexlab\endcsname\relax\def\natexlab#1{#1}\fi
\providecommand{\url}[1]{\href{#1}{#1}}
\providecommand{\dodoi}[1]{doi:~\href{http://doi.org/#1}{\nolinkurl{#1}}}
\providecommand{\doeprint}[1]{\href{http://ascl.net/#1}{\nolinkurl{http://ascl.net/#1}}}
\providecommand{\doarXiv}[1]{\href{https://arxiv.org/abs/#1}{\nolinkurl{https://arxiv.org/abs/#1}}}

\bibitem[{{Abhishek} {et~al.}(2022){Abhishek}, {Malusare}, {Tanushree}, {Hegde}, \& {Konar}}]{Abhishek22}
{Abhishek}, {Malusare}, N., {Tanushree}, N., {Hegde}, G., \& {Konar}, S. 2022, Journal of Astrophysics and Astronomy, 43, 75, \dodoi{10.1007/s12036-022-09862-3}

\bibitem[{{Bera} \& {Chengalur}(2019)}]{Bera19}
{Bera}, A., \& {Chengalur}, J.~N. 2019, \mnras, 490, L12, \dodoi{10.1093/mnrasl/slz140}

\bibitem[{{Bhattacharyya} {et~al.}(2018){Bhattacharyya}, {Lyne}, {Stappers}, {Weltevrede}, {Keane}, {McLaughlin}, {Kramer}, {Jordan}, \& {Bassa}}]{Bhattacharyya18}
{Bhattacharyya}, B., {Lyne}, A.~G., {Stappers}, B.~W., {et~al.} 2018, \mnras, 477, 4090, \dodoi{10.1093/mnras/sty923}

\bibitem[{{Biggs}(1992)}]{Biggs92}
{Biggs}, J.~D. 1992, \apj, 394, 574, \dodoi{10.1086/171608}

\bibitem[{{Burke-Spolaor}(2013)}]{Burke-Spolaor13}
{Burke-Spolaor}, S. 2013, in Neutron Stars and Pulsars: Challenges and Opportunities after 80 years, ed. J.~{van Leeuwen}, Vol. 291, 95--100, \dodoi{10.1017/S1743921312023277}

\bibitem[{{Burke-Spolaor} \& {Bailes}(2010)}]{Burke-Spolaor10}
{Burke-Spolaor}, S., \& {Bailes}, M. 2010, \mnras, 402, 855, \dodoi{10.1111/j.1365-2966.2009.15965.x}

\bibitem[{{Cordes} \& {Lazio}(2002)}]{ne2001}
{Cordes}, J.~M., \& {Lazio}, T.~J.~W. 2002, arXiv e-prints, astro, \dodoi{10.48550/arXiv.astro-ph/0207156}

\bibitem[{{Cui} {et~al.}(2017){Cui}, {Boyles}, {McLaughlin}, \& {Palliyaguru}}]{Cui17}
{Cui}, B.~Y., {Boyles}, J., {McLaughlin}, M.~A., \& {Palliyaguru}, N. 2017, \apj, 840, 5, \dodoi{10.3847/1538-4357/aa6aa9}

\bibitem[{{Deneva} {et~al.}(2009){Deneva}, {Cordes}, {McLaughlin}, {Nice}, {Lorimer}, {Crawford}, {Bhat}, {Camilo}, {Champion}, {Freire}, {Edel}, {Kondratiev}, {Hessels}, {Jenet}, {Kasian}, {Kaspi}, {Kramer}, {Lazarus}, {Ransom}, {Stairs}, {Stappers}, {van Leeuwen}, {Brazier}, {Venkataraman}, {Zollweg}, \& {Bogdanov}}]{Deneva09}
{Deneva}, J.~S., {Cordes}, J.~M., {McLaughlin}, M.~A., {et~al.} 2009, \apj, 703, 2259, \dodoi{10.1088/0004-637X/703/2/2259}

\bibitem[{{Enoto} {et~al.}(2019){Enoto}, {Kisaka}, \& {Shibata}}]{Enoto19}
{Enoto}, T., {Kisaka}, S., \& {Shibata}, S. 2019, Reports on Progress in Physics, 82, 106901, \dodoi{10.1088/1361-6633/ab3def}

\bibitem[{{Everett} \& {Weisberg}(2001)}]{Everett01}
{Everett}, J.~E., \& {Weisberg}, J.~M. 2001, \apj, 553, 341, \dodoi{10.1086/320652}

\bibitem[{{Geyer} {et~al.}(2021){Geyer}, {Serylak}, {Abbate}, {Bailes}, {Buchner}, {Chilufya}, {Johnston}, {Karastergiou}, {Main}, {van Straten}, \& {Shamohammadi}}]{Geyer21}
{Geyer}, M., {Serylak}, M., {Abbate}, F., {et~al.} 2021, \mnras, 505, 4468, \dodoi{10.1093/mnras/stab1501}

\bibitem[{{Gil} {et~al.}(2007){Gil}, {Melikidze}, \& {Zhang}}]{Gil07}
{Gil}, J., {Melikidze}, G., \& {Zhang}, B. 2007, \mnras, 376, L67, \dodoi{10.1111/j.1745-3933.2007.00288.x}

\bibitem[{{Gil} \& {Sendyk}(2000)}]{Gil00}
{Gil}, J.~A., \& {Sendyk}, M. 2000, \apj, 541, 351, \dodoi{10.1086/309394}

\bibitem[{{Goldreich} \& {Julian}(1969)}]{Goldreich69}
{Goldreich}, P., \& {Julian}, W.~H. 1969, \apj, 157, 869, \dodoi{10.1086/150119}

\bibitem[{{Hickish} {et~al.}(2016){Hickish}, {Abdurashidova}, {Ali}, {Buch}, {Chaudhari}, {Chen}, {Dexter}, {Domagalski}, {Ford}, {Foster}, {George}, {Greenberg}, {Greenhill}, {Isaacson}, {Jiang}, {Jones}, {Kapp}, {Kriel}, {Lacasse}, {Lutomirski}, {MacMahon}, {Manley}, {Martens}, {McCullough}, {Muley}, {New}, {Parsons}, {Price}, {Primiani}, {Ray}, {Siemion}, {van Tonder}, {Vertatschitsch}, {Wagner}, {Weintroub}, \& {Werthimer}}]{Hickish16}
{Hickish}, J., {Abdurashidova}, Z., {Ali}, Z., {et~al.} 2016, Journal of Astronomical Instrumentation, 5, 1641001, \dodoi{10.1142/S2251171716410014}

\bibitem[{{Hotan} {et~al.}(2004){Hotan}, {van Straten}, \& {Manchester}}]{Hotan04}
{Hotan}, A.~W., {van Straten}, W., \& {Manchester}, R.~N. 2004, \pasa, 21, 302, \dodoi{10.1071/AS04022}

\bibitem[{{Jiang} {et~al.}(2019){Jiang}, {Yue}, {Gan}, {Yao}, {Li}, {Pan}, {Sun}, {Yu}, {Liu}, {Tang}, {Qian}, {Lu}, {Yan}, {Peng}, {Zhang}, {Wang}, {Li}, \& {Li}}]{Jiang19}
{Jiang}, P., {Yue}, Y., {Gan}, H., {et~al.} 2019, Science China Physics, Mechanics, and Astronomy, 62, 959502, \dodoi{10.1007/s11433-018-9376-1}

\bibitem[{{Jiang} {et~al.}(2020){Jiang}, {Tang}, {Hou}, {Liu}, {Kr{\v{c}}o}, {Qian}, {Sun}, {Ching}, {Liu}, {Duan}, {Yue}, {Gan}, {Yao}, {Li}, {Pan}, {Yu}, {Liu}, {Li}, {Peng}, {Yan}, \& {FAST Collaboration}}]{Jiang20}
{Jiang}, P., {Tang}, N.-Y., {Hou}, L.-G., {et~al.} 2020, Research in Astronomy and Astrophysics, 20, 064, \dodoi{10.1088/1674-4527/20/5/64}

\bibitem[{{Keane} {et~al.}(2011){Keane}, {Kramer}, {Lyne}, {Stappers}, \& {McLaughlin}}]{Keane11}
{Keane}, E.~F., {Kramer}, M., {Lyne}, A.~G., {Stappers}, B.~W., \& {McLaughlin}, M.~A. 2011, \mnras, 415, 3065, \dodoi{10.1111/j.1365-2966.2011.18917.x}

\bibitem[{{Kuzmin}(2007)}]{Kuzmin07}
{Kuzmin}, A.~D. 2007, \apss, 308, 563, \dodoi{10.1007/s10509-007-9347-5}

\bibitem[{{Lorimer} \& {Kramer}(2004)}]{Lorimer04}
{Lorimer}, D.~R., \& {Kramer}, M. 2004, {Handbook of Pulsar Astronomy}, Vol.~4

\bibitem[{{Manchester} {et~al.}(2005){Manchester}, {Hobbs}, {Teoh}, \& {Hobbs}}]{Manchester05}
{Manchester}, R.~N., {Hobbs}, G.~B., {Teoh}, A., \& {Hobbs}, M. 2005, \aj, 129, 1993, \dodoi{10.1086/428488}

\bibitem[{{Manchester} {et~al.}(1975){Manchester}, {Taylor}, \& {Huguenin}}]{Manchester75}
{Manchester}, R.~N., {Taylor}, J.~H., \& {Huguenin}, G.~R. 1975, \apj, 196, 83, \dodoi{10.1086/153395}

\bibitem[{{McKinnon} \& {Stinebring}(1998)}]{McKinnon98}
{McKinnon}, M.~M., \& {Stinebring}, D.~R. 1998, \apj, 502, 883, \dodoi{10.1086/305924}

\bibitem[{{McLaughlin} {et~al.}(2006){McLaughlin}, {Lyne}, {Lorimer}, {Kramer}, {Faulkner}, {Manchester}, {Cordes}, {Camilo}, {Possenti}, {Stairs}, {Hobbs}, {D'Amico}, {Burgay}, \& {O'Brien}}]{McLaughlin06}
{McLaughlin}, M.~A., {Lyne}, A.~G., {Lorimer}, D.~R., {et~al.} 2006, \nat, 439, 817, \dodoi{10.1038/nature04440}

\bibitem[{{McLaughlin} {et~al.}(2009){McLaughlin}, {Lyne}, {Keane}, {Kramer}, {Miller}, {Lorimer}, {Manchester}, {Camilo}, \& {Stairs}}]{McLaughlin09}
{McLaughlin}, M.~A., {Lyne}, A.~G., {Keane}, E.~F., {et~al.} 2009, \mnras, 400, 1431, \dodoi{10.1111/j.1365-2966.2009.15584.x}

\bibitem[{{Medin} \& {Lai}(2007)}]{Medin07}
{Medin}, Z., \& {Lai}, D. 2007, \mnras, 382, 1833, \dodoi{10.1111/j.1365-2966.2007.12492.x}

\bibitem[{{Popov} {et~al.}(2006){Popov}, {Soglasnov}, {Kondrat'Ev}, {Kostyuk}, {Ilyasov}, \& {Oreshko}}]{Popov06}
{Popov}, M.~V., {Soglasnov}, V.~A., {Kondrat'Ev}, V.~I., {et~al.} 2006, Astronomy Reports, 50, 55, \dodoi{10.1134/S1063772906010069}

\bibitem[{{Radhakrishnan} \& {Cooke}(1969)}]{Radhakrishnan69}
{Radhakrishnan}, V., \& {Cooke}, D.~J. 1969, \aplett, 3, 225

\bibitem[{{Ruderman} \& {Sutherland}(1975)}]{Ruderman75}
{Ruderman}, M.~A., \& {Sutherland}, P.~G. 1975, \apj, 196, 51, \dodoi{10.1086/153393}

\bibitem[{{Szary}(2013)}]{Szary13}
{Szary}, A. 2013, arXiv e-prints, arXiv:1304.4203, \dodoi{10.48550/arXiv.1304.4203}

\bibitem[{{Szary} {et~al.}(2015{\natexlab{a}}){Szary}, {Melikidze}, \& {Gil}}]{Szary15a}
{Szary}, A., {Melikidze}, G.~I., \& {Gil}, J. 2015{\natexlab{a}}, \mnras, 447, 2295, \dodoi{10.1093/mnras/stu2622}

\bibitem[{{Szary} {et~al.}(2015{\natexlab{b}}){Szary}, {Melikidze}, \& {Gil}}]{Szary15b}
---. 2015{\natexlab{b}}, \apj, 800, 76, \dodoi{10.1088/0004-637X/800/1/76}

\bibitem[{{Timokhin}(2007)}]{Timokhin07}
{Timokhin}, A.~N. 2007, \mnras, 379, 605, \dodoi{10.1111/j.1365-2966.2007.11864.x}

\bibitem[{{van Straten} \& {Bailes}(2011)}]{Straten11}
{van Straten}, W., \& {Bailes}, M. 2011, \pasa, 28, 1, \dodoi{10.1071/AS10021}

\bibitem[{{Wang} {et~al.}(2007){Wang}, {Manchester}, \& {Johnston}}]{Wang07}
{Wang}, N., {Manchester}, R.~N., \& {Johnston}, S. 2007, \mnras, 377, 1383, \dodoi{10.1111/j.1365-2966.2007.11703.x}

\bibitem[{{Wang} {et~al.}(2019){Wang}, {Lu}, {Zhang}, {Chen}, {Luo}, \& {Xu}}]{Wang19}
{Wang}, W., {Lu}, J., {Zhang}, S., {et~al.} 2019, Science China Physics, Mechanics, and Astronomy, 62, 979511, \dodoi{10.1007/s11433-018-9334-y}

\bibitem[{{Weltevrede} {et~al.}(2006){Weltevrede}, {Stappers}, {Rankin}, \& {Wright}}]{Weltevrede06}
{Weltevrede}, P., {Stappers}, B.~W., {Rankin}, J.~M., \& {Wright}, G.~A.~E. 2006, \apjl, 645, L149, \dodoi{10.1086/506346}

\bibitem[{{Yao} {et~al.}(2017){Yao}, {Manchester}, \& {Wang}}]{YMW16}
{Yao}, J.~M., {Manchester}, R.~N., \& {Wang}, N. 2017, \apj, 835, 29, \dodoi{10.3847/1538-4357/835/1/29}

\bibitem[{{Zhang} {et~al.}(2024){Zhang}, {Geng}, {Wang}, {Yang}, {Kaczmarek}, {Tang}, {Johnston}, {Hobbs}, {Manchester}, {Wu}, {Jiang}, {Huang}, {Zou}, {Dai}, {Zhang}, {Li}, {Yang}, {Dai}, {Chang}, {Pan}, {Lu}, {Wei}, {Li}, {Wu}, {Qian}, {Wang}, {Wang}, {Feng}, \& {Staveley-Smith}}]{Zhang23}
{Zhang}, S.~B., {Geng}, J.~J., {Wang}, J.~S., {et~al.} 2024, \apj, 972, 59, \dodoi{10.3847/1538-4357/ad6602}

\bibitem[{{Zhou} {et~al.}(2023){Zhou}, {Han}, {Xu}, {Wang}, {Wang}, {Wang}, {Jing}, {Chen}, {Yan}, {Su}, {Gan}, {Jiang}, {Sun}, {Wang}, {Wang}, {Wang}, {Xu}, \& {You}}]{Zhou23}
{Zhou}, D.~J., {Han}, J.~L., {Xu}, J., {et~al.} 2023, Research in Astronomy and Astrophysics, 23, 104001, \dodoi{10.1088/1674-4527/accc76}

\end{thebibliography}
\bibliographystyle{aasjournal}
\clearpage



\end{document}